# Unification of the wave and guidance equations for spin $\frac{1}{2}$

Peter Holland

Green Templeton College, University of Oxford, Oxford, UK

peter.holland@gtc.ox.ac.uk

## Abstract

We generalize our previous unification of the Schrödinger and guidance equations in a single inhomogeneous Schrödinger equation to a Riemannian manifold with an external vector potential. A special case yields the unified theory for a spin $\frac{1}{2}$ rigid rotator. The theory is proved to be symmetrical under the Galileo group, the unified field that integrates the particle and guiding wave being a 2-spinor.

## 1 Introduction

In previous work ([1] and references therein), we examined a composite system comprising a wave and a particle whose collective state, embodied in a field $u(x, t)$, obeys the inhomogeneous Schrödinger equation

$$i\hbar \frac{\partial u(x,t)}{\partial t} = \left(-\frac{\hbar^2}{2m}\partial_{ii} + V\right)u(x,t) + 2\frac{\delta Q(\rho(q))}{\delta \psi^*(x)}\bigg|_{q=q(t,q_0)} \quad (1)$$

where $\psi = \sqrt{\rho}e^{iS/\hbar}$ satisfies the homogeneous (Schrödinger) equation, $Q$ is the quantum potential constructed from $\psi$, and $q_i(q_0, t)$ are the coordinates of the particle with initial position $q_{0i}$, $i = 1,2,3$. A fundamental property of the inhomogeneous equation is that the solution

$$u(x, t, q_0) = \psi(x, t) - \frac{1}{\psi^*(x,t)}\delta(x - q(t, q_0)) \quad (2)$$

renders it equivalent to the de Broglie-Bohm guidance equation for the particle, $\dot{q}_i = m^{-1}\partial_i S(x,t)|_{x=q(t,q_0)}$. This result emerged from an analytical formulation of the wave-particle interaction. It was shown that the approach provides an alternative basis for the de Broglie-Bohm theory, and resolves several problems with the latter's conventional presentation: it detaches the justification for the guidance law from statistics; it incorporates the lack of reaction of the particle on the $\psi$-wave within a general theoretical framework; and it integrates the inseparable yet disparate wave and particle elements in a single field $u$, the $\psi$-wave being its sourceless homogeneous component while the (point) particle is represented by a highly concentrated solitonic amplitude (the delta function) that moves in accordance with the guidance law. Correspondingly, the wave and guidance equations are amalgamated in the single inhomogeneous equation (1), to whose source the particle contributes via the quantum potential. This melding of wave and particle equations readily extends to a many-body

system. Further details of the background and application of the approach are given in [1].

In this contribution, we broaden the model to include (non-relativistic) spin $\frac{1}{2}$. In this quest one cannot simply replace $\psi$ in (2) by a spinor field, which, due to the appearance of $1/\psi$, would not give a mathematically meaningful expression. Our method is to first generalize the previous results to an *N*-dimensional Riemannian manifold $\mathfrak{M}$ with an external vector potential. This is done using a different technique to the analytical formulation of [1]. Following the de Broglie-Bohm prescription, the particle 'hidden' variables are positions represented by generalized coordinates in the *N*-space. The method engenders unified theories for a wide class of systems governed by the generalized Schrödinger equation, distinguished by the choice of $\mathfrak{M}$ and the states thereon.

There are two aspects of the generalized treatment that did not feature in [1] but do in the application to spin. The first is that the configuration space $\widetilde{\mathfrak{M}}$ of the particle need not coincide with $\mathfrak{M}$; the latter may involve a group manifold, for example. The second aspect concerns how the particle component of the composite is represented by the unified field. In the case of a point particle, a non-spreading delta function whose dynamics is governed by the guidance law (as in (2)) renders the physical system faithfully. In the more general context of a particle with structure, the delta function may not provide a full field-theoretic model, even though it follows the system trajectory.

The spin $\frac{1}{2}$ case is characterized by a subspace of states on the manifold $\mathfrak{M} = \mathbb{R}^3 \otimes SU(2)$, and our approach naturally generates a unified theory for a spherical rigid rotator in the angular coordinate representation. In this formalism the unified field may be expressed consistently in a form similar to (2), while the particle position variables comprise translational and rotational freedoms. The rotator theory is pertinent to problems where moment of inertia plays a structural role, such as the rotational spectra of nuclei [2]. The causal interpretation of the spin $\frac{1}{2}$ rotator was developed and proved to be consistent with quantum predictions in [3-5] (and applied, in particular, to the Stern-Gerlach and EPR experiments; for further applications, see [6]). Similar techniques were used previously in the stochastic interpretation [7].

We shall show that the theory is symmetrical with respect to the Galileo group, the unified field being a 2-spinor in the angular momentum (spinor) representation. While it may be couched in spinorial terms, the theory should be distinguished from the de Broglie-Bohm approach to spin that employs just translational hidden variables, which is tied to the spinor representation of the underlying quantum theory [5]. It is uncertain whether a unified model for spin of the type we are investigating can be developed using just the translational freedom. We have described previously how the causal theory based on the spinor representation, although consistent with quantum predictions, has shortcomings that are considerably alleviated by employing the angular coordinate representation [3,5].

## 2 Unification of the Schrödinger and guidance equations in Riemannian space

Consider an $N$-dimensional Riemannian manifold $\mathfrak{M}$ equipped with generalized coordinates $x^\mu$ and (static) metric $g_{\mu\nu}(x)$ ($\mu, \nu = 1, \ldots, N$), defined so that the line element $\sqrt{g_{\mu\nu}dx^\mu dx^\nu}$ has the dimension of length. Define $g = |\det g_{\mu\nu}|$. Denoting prescribed external vector and scalar potentials by $A_\mu(x,t)$ and $V(x,t)$, respectively, the Schrödinger equation

$$i\hbar \frac{\partial \psi}{\partial t} = -\frac{\hbar^2}{2m\sqrt{g}}(\partial_\mu - (i/\hbar)A_\mu)[\sqrt{g}g^{\mu\nu}(\partial_\nu \psi - (i/\hbar)A_\nu \psi)] + V\psi, \tag{3}$$

is covariant under general point transformations $x^\mu \to x'^\mu(x)$. We assume that the (scalar) wavefunction $\psi(x,t)$ is normalized with respect to the measure $\sqrt{g}d^N x$:

$$\int |\psi(x,t)|^2 \sqrt{g}d^N x = 1. \tag{4}$$

The system has mass $m$ and, as we shall see in Sect. 3, the theory embraces systems with structure. Within this general scheme our first task *en route* to a unified theory of particle and wave is to show that (3) may be modified so that the resulting equation is equivalent to a continuity equation in $\mathfrak{M}$. It is the latter property that establishes the connection with the guidance law.

Let a function $u(x,t)$ denote the state of a continuous system and consider the function obtained by applying the 'Schrödinger operator' to it:

$$F(u) \equiv \left[i\hbar \frac{\partial}{\partial t} + \frac{\hbar^2}{2m\sqrt{g}}(\partial_\mu - (i/\hbar)A_\mu)[\sqrt{g}g^{\mu\nu}(\partial_\nu - (i/\hbar)A_\nu)] - V\right]u. \tag{5}$$

Evidently, the equation $F(\psi) = 0$ coincides with (3). We are interested in $F$ when $u$ deviates from $\psi$. Specifically, suppose $u = \psi - f/\psi^*$ where $\psi$ obeys (3) and $f(x,t)$ is an unknown complex scalar function whose interpretation is to be determined. A straightforward calculation shows that

$$F(u) = -\frac{i\hbar}{\psi^*}\left[\frac{\partial f}{\partial t} + \frac{1}{\sqrt{g}}\partial_\mu(f\sqrt{g}v^\mu)\right] + G(\psi, f) \tag{6}$$

where

$$v^\mu(x,t) = m^{-1}g^{\mu\nu}(\partial_\nu S - A_\nu) \tag{7}$$

is a vector field and

$$G(x,t) = \frac{\hbar^2}{2m\sqrt{g}\psi^*}[f\partial_\mu(\sqrt{g}g^{\mu\nu}\partial_\nu \log\rho) + \sqrt{g}g^{\mu\nu}\partial_\mu f\partial_\nu \log\rho - \partial_\mu(\sqrt{g}g^{\mu\nu}\partial_\nu f)] \tag{8}$$

is a scalar field. Suppose we set $F = G$. Then $u$ obeys an inhomogeneous Schrödinger equation whose 'source' $G$ is built from the solution $\psi$ of the homogeneous equation (3) and the function $f$. To find $f$, we observe from (6) that the modified Schrödinger equation $F = G$ with the solution $u = \psi - f/\psi^*$ is equivalent to a continuity equation,

$$\frac{\partial(f\sqrt{g})}{\partial t} + \partial_\mu(f\sqrt{g}v^\mu) = 0, \tag{9}$$

as we set out to prove. Identifying $v^\mu$ as a velocity field, the interpretation of $f$ is, therefore, that $f\sqrt{g}$ is a (in general, complex) scalar density conserved by the flow represented by $v^\mu$. The function $u$ is, therefore, fully determined once $\psi$ and $f(x, t = 0)$ are specified. Note that, with the choice $f = |\psi|^2$ (i.e., $u = 0$), (9) coincides with the continuity equation implied by the homogeneous equation (3).

We come now to the case of interest. Suppose that the $\psi$ wave is accompanied by a physical system (a 'particle', which may comprise many three-dimensional particles having internal freedoms) that traces out a one-dimensional track in a configuration space $\widetilde{\mathfrak{M}}$ whose coordinates $q^\mu(q_0, t)$ correspond to the arguments $x^\mu$ of the wavefunction, where $q_0^\mu$ specifies the initial position. The additional system's microscopic density is an $N$-dimensional delta function $\delta(x - q)$, an entity that retains its integrity whilst traversing the particle track. For a suitably chosen velocity field, the governing dynamical law of the microdensity is a continuity equation since, as we show in a moment, this is equivalent to the particle law. Pursuing this idea, we identify the density with the conserved function $f\sqrt{g}$, that is, we choose $f(x, t) = g(x)^{-1/2}\delta(x - q(t))$, which is a scalar with respect to arbitrary transformations of $x$ [8]. The function $u$ is, therefore, taken to be the scalar field

$$u(x, t, q_0) = \psi(x, t) - \frac{1}{\sqrt{g}\psi^*(x, t)}\delta(x - q(q_0, t)). \tag{10}$$

We shall demonstrate two key properties of this model. The first is that *the inhomogeneous equation $F = G$ with solution (10) is equivalent to the de Broglie-Bohm guidance equation*. This is easily proved via the medium of the continuity equation (9), which, as we have shown, is equivalent to $F = G$ with solution (10). Inserting $f\sqrt{g} = \delta(x - q)$ in (9) and subtracting the identity

$$\frac{\partial \delta(x - q)}{\partial t} + \dot{q}^\mu \partial_\mu \delta(x - q) = 0 \tag{11}$$

implies that $\big(\dot{q}^\mu - v^\mu(x = q(q_0, t), t)\big)\partial_\mu \delta(x - q) = 0$. Multiplying the latter relation by an arbitrary function of $x^\mu$ and integrating over $x^\mu$, we deduce the de Broglie-Bohm law

$$\dot{q}^\mu = m^{-1}g^{\mu\nu}(\partial_\nu S - A_\nu)|_{x=q(t,q_0)}. \tag{12}$$

Conversely, we can deduce from (11) and (12) that the microdensity obeys (9), and hence that (10) satisfies $F = G$. The trajectory conserves the local probability $|\psi(q(t), t)|^2\sqrt{g}d^N q(t)$.

The second property is that *the source term (8) derives from the functional derivative of the quantum potential*

$$Q(\rho(x)) = -\frac{\hbar^2}{2m\sqrt{g\rho}}\partial_\mu(\sqrt{g}g^{\mu\nu}\partial_\nu\sqrt{\rho}). \tag{13}$$

Evaluating $Q$ along the trajectory, its functional derivative with respect to $\psi^*(x)$ is given by $\delta Q(\rho(q))/\delta\psi^*(x) = \psi\,\delta Q(q)/\delta\rho(x)$, where

$$\frac{\delta Q(q)}{\delta\rho(x)} = \frac{\partial Q(x)}{\partial\rho(x)}\delta(x-q) - \partial_\mu\left[\frac{\partial Q}{\partial(\partial_\mu\rho)}\delta(x-q)\right] + \partial_{\mu\nu}\left[\frac{\partial Q}{\partial(\partial_{\mu\nu}\rho)}\delta(x-q)\right]. \tag{14}$$

To evaluate this expression, we write $Q\delta(x-q)$ as $\sqrt{g}Qf$. Then, using (13) and referring to (8), we find

$$\frac{2}{\sqrt{g}}\frac{\delta Q(q)}{\delta\psi^*(x)} = G\left(\psi, f = g^{-1/2}\delta(x-q)\right). \tag{15}$$

Combining these results, we have proved the following generalization of our previous result [1]. Let a material system of mass $m$ be associated with a complex field $u(x,t,q_0)$, which obeys the following inhomogeneous Schrödinger equation in $\mathfrak{M}$,

$$i\hbar\frac{\partial u}{\partial t} = -\frac{\hbar^2}{2m\sqrt{g}}(\partial_\mu - (i/\hbar)A_\mu)\left[\sqrt{g}g^{\mu\nu}(\partial_\nu u - (i/\hbar)A_\nu u)\right] + Vu + \frac{2}{\sqrt{g}}\frac{\delta Q(\rho(q))}{\delta\psi^*(x)}, \tag{16}$$

where $\psi$ satisfies the homogeneous (Schrödinger) equation, $Q$ is the quantum potential (13) constructed from $\psi$, $q^\mu(q_0,t)$ are the coordinates of a particle with initial position $q_0^\mu$, and the source term is given by (8) and (15). Then the function (10) satisfies (16) if and only if the particle coordinates obey the guidance formula (12).

A feature of the model is that it incorporates an account of the passage to the classical limit whilst retaining the same basic structure, insofar as this limit can ever be attained successfully. Thus, if the relative values of the quantum potential and force are negligible compared with the classical energy and force in a certain spacetime region, the particle motion there will be classical-like. In this regime, the structure (10) remains intact but the particle is guided solely by classical potentials (the limiting process is applied after taking the functional derivative in (16)). The subtle issues involved in this inter-theory liminal domain (see [9] and references therein) persist in the unified theory but the latter does not introduce additional complications.

### 3 Deduction of the spin $\frac{1}{2}$ Schrödinger equation

To show how the wave and guidance equations for a spin $\frac{1}{2}$ system may be united in an inhomogeneous wave equation, we specialize the preceding treatment. The first task is to obtain the spin $\frac{1}{2}$ Schrödinger equation. Consider the six-dimensional manifold $\mathfrak{M} = \mathbb{R}^3 \otimes SU(2)$ with coordinates $x^\mu = (x_i, \alpha^r)$, where $i,j,...$ and $r,s,... = 1,2,3$, and $\alpha^r = (\alpha,\beta,\gamma)$ are Euler angles with $\alpha \in [0,\pi]$, $\beta \in [0,2\pi]$, $\gamma \in [0,4\pi]$ (for background see [5]). The metric on $\mathfrak{M}$ is given by [10]

$$g_{\mu\nu} = \begin{pmatrix} \delta_{ij} & 0 \\ 0 & g_{rs} \end{pmatrix}, \quad g_{rs} = l^2 \begin{pmatrix} 1 & 0 & 0 \\ 0 & 1 & \cos\alpha \\ 0 & \cos\alpha & 1 \end{pmatrix},$$
$$g^{rs} = l^{-2} \begin{pmatrix} 1 & 0 & 0 \\ 0 & \operatorname{cosec}^2\alpha & -\cot\alpha\,\operatorname{cosec}\alpha \\ 0 & -\cot\alpha\,\operatorname{cosec}\alpha & \operatorname{cosec}^2\alpha \end{pmatrix},$$
(17)

where $l$ is a constant with the dimension of length and $g = l^6 \sin^2\alpha$. We shall restrict the Hilbert space of states on $\mathfrak{M}$ to the spin $\frac{1}{2}$ subspace. In the angular coordinate representation, the wavefunction is written

$$\psi(x, \alpha, t) = \Psi^a(x, t) u_a(\alpha), \quad a = 1, 2, \tag{18}$$

where the Pauli 2-spinor field $\Psi^a$ is the state in the angular momentum representation and the basis (or transformation) functions $u_a(\alpha)$ are

$$u_1 = (2\sqrt{2}\pi)^{-1}\cos(\alpha/2)e^{-i(\beta+\gamma)/2}, \quad u_2 = -i(2\sqrt{2}\pi)^{-1}\sin(\alpha/2)e^{i(\beta-\gamma)/2}. \tag{19}$$

The latter obey the orthonormality condition

$$\int u_a^*(\alpha)\, u_b(\alpha) d\Omega = \delta_b^a, \quad d\Omega = \sin\alpha\, d\alpha\, d\beta\, d\gamma, \tag{20}$$

from which the inverse of (18) follows:

$$\Psi^a(x, t) = \int u_a^*(\alpha)\, \psi(x, \alpha, t) d\Omega. \tag{21}$$

It is convenient to represent differentiation with respect to the angles via the angular momentum operators

$$\widehat{M}_i = -i\hbar A_i^r \partial_r, \tag{22}$$

where $\partial_r = \partial/\partial\alpha^r$ and

$$A_i^r = \begin{pmatrix} -\cos\beta & \sin\beta \cot\alpha & -\sin\beta \operatorname{cosec}\alpha \\ \sin\beta & \cos\beta \cot\alpha & -\cos\beta \operatorname{cosec}\alpha \\ 0 & -1 & 0 \end{pmatrix}. \tag{23}$$

The operators $\widehat{M}_i$ obey the angular momentum exchange relations and, when applied to the function (18), those of a Clifford algebra:

$$[\widehat{M}_i, \widehat{M}_j] = i\hbar\varepsilon_{ijk}\widehat{M}_k, \quad \widehat{M}_i\widehat{M}_j + \widehat{M}_j\widehat{M}_i = 2(\hbar/2)^2\delta_{ij}, \quad i,j,k = 1,2,3. \tag{24}$$

The matrix (23) satisfies the following differential identity

$$\partial_r(\sin\alpha\, A_i^r) = 0 \tag{25}$$

and is connected to the metric via the relation

$$g^{rs} = l^{-2} A_i^r A_i^s. \tag{26}$$

Writing the external potentials $A^\mu = \big(A^i(x,t), A^r(x,\alpha,t)\big)$, inserting (17), and using (22), (25) and (26), the generalized Schrödinger equation (3) becomes

$$i\hbar \frac{\partial \psi}{\partial t} = -\frac{\hbar^2}{2m}(\partial_i - (i/\hbar)A_i)^2 \psi + \frac{1}{2I}\left(\widehat{M}_i + \mathfrak{m}IB_i\right)^2 \psi + V\psi \qquad (27)$$

with wavefunction (18). Here $I = ml^2$ and we have chosen $A_r(x,\alpha,t)$ so that $A_i^r A_r = -\mathfrak{m}IB_i(x,t)$ where $B_i = (\nabla \times \boldsymbol{A})_i$. We thus obtain from our generalized treatment the Schrödinger ('Bopp-Haag' [11]) equation for the translational and rotational motion of a spherical spin $\tfrac{1}{2}$ rigid rotator of mass $m$, moment of inertia $I$, charge unity and magnetic moment $\mathfrak{m}$ in external magnetic and scalar potentials.

Eq. (27) may be simplified by observing that $\widehat{M}_i^2 = 3\hbar^2/4$ when applied to a spin $\tfrac{1}{2}$ state, and that the associated term $3\hbar^2 \psi/8I$ in (27) may be absorbed in the global phase. If we suppose in addition that the term $I(\mathfrak{m}B_i)^2/2$ may be neglected compared to other relevant energies (say, by assuming $B_i$ is weak[1]), the wave equation (27) becomes

$$i\hbar \frac{\partial \psi}{\partial t} = -\frac{\hbar^2}{2m}(\partial_i - (i/\hbar)A_i)^2 \psi + \mathfrak{m}B_i \widehat{M}_i \psi + V\psi. \qquad (28)$$

Multiplying (28) by $\sin\alpha \, u_a^*$ and using (21) gives

$$i\hbar \frac{\partial \Psi^a}{\partial t} = \left[-\frac{\hbar^2}{2m}(\partial_i - (i/\hbar)A_i)^2 + V\right]\Psi^a + \mathfrak{m}B_i \sigma_{ib}^a \Psi^b \qquad (29)$$

where $\sigma_{ib}^a = \int u_a^* \widehat{M}_i u_b d\Omega$ are the Pauli matrices

$$\sigma_1 = \begin{pmatrix} 0 & 1 \\ 1 & 0 \end{pmatrix}, \quad \sigma_2 = \begin{pmatrix} 0 & -i \\ i & 0 \end{pmatrix}, \quad \sigma_3 = \begin{pmatrix} 1 & 0 \\ 0 & -1 \end{pmatrix}. \qquad (30)$$

We thus obtain the Pauli equation (29) as a special case of our general treatment, (28) being its angular coordinate version. For generality, we continue to work with (27).

Note that the spatial probability density is obtained from the same formal procedure in each of the spinor and angular formulations, summation over the spin indices: $\Psi^{a*}\Psi^a = \int |\psi|^2 d\Omega$.

## 4 Unified spin $\tfrac{1}{2}$ field

The unified theory for spin $\tfrac{1}{2}$ may now be written down immediately. First, as in Sect. 2, we introduce a trajectory whose coordinates correspond to the arguments of the wavefunction, which are here three translation and three rotation coordinates: $q^\mu(t) = (q_i(t), \theta^r(t))$ with initial values $q_0^\mu = (q_{0i}, \theta_0^r)$. The velocity field (7) splits up into translational ($v_i$) and angular ($v^r$) components, and the guidance equation (12) becomes the six coupled relations

$$\dot{q}_i = v_i = m^{-1}(\partial_i S - A_i)|_{x=q(t,q_0,\theta_0),\alpha=\theta(t,q_0,\theta_0)} \quad (31)$$

$$\dot{\theta}^r = v^r = I^{-1}A_i^r\big((-i\hbar)^{-1}\widehat{M}_i S + \mathfrak{m}IB_i\big)\big|_{x=q(t,q_0,\theta_0),\alpha=\theta(t,q_0,\theta_0)}. \quad (32)$$

---

[1] We may obtain (28) from (27) by passing to the limit $l \to 0$ of a point particle (in such a way that the mass, charge and magnetic moment remain finite [5]). This procedure is desirable if we wish to avoid committing to a particular model of a structured particle. However, we do not include this option here because the guidance formula (32) involves $I$.

Writing the solution (18) as $\psi = \sqrt{\rho}e^{iS/\hbar}$, these are the trajectory equations in the de Broglie-Bohm theory of a spin $\frac{1}{2}$ rigid rotator, where $q_i(t)$ denotes the centre of mass coordinates and $\theta^r(t)$ parameterizes the orientation (defined by a rotation matrix) of a frame fixed in the body [5]. The angular momentum (spin) vector of the body is $IA_r^i\dot{\theta}^r$. The configuration space of the orbits is that of a classical rigid rotator, $\widetilde{\mathfrak{M}} = \mathbb{R}^3 \otimes SO(3)$, and the local probability $|\psi(q(t),\theta(t))|^2 d^3q(t)d\Omega(\theta(t))$ is conserved along each. Note that we can apply the theory to the Pauli equation (28) but $I$ in (32) is then a free parameter.

Next, to obtain the unified formalism, we write the inhomogeneous equation and unified field in terms of redefined amplitudes $\tilde{\psi} = l^{3/2}\psi$, $\widetilde{\Psi}^a = l^{3/2}\Psi^a$ and $\tilde{u} = l^{3/2}u$. Then the normalization condition (4) becomes the usual one for a spinor field: $\int \tilde{\psi}^*\tilde{\psi} d^3x d\Omega = \int \widetilde{\Psi}^\dagger \widetilde{\Psi} d^3x = 1$. Noting that $Q(\tilde{\rho}) = Q(\rho)$ and $\delta/\delta\tilde{\psi}^* = l^{-3/2}\delta/\delta\psi^*$, the inhomogeneous spin $\frac{1}{2}$ equation obtained as a special case of (16) becomes, in terms of the redefined variables (we henceforth drop the tildes),

$$i\hbar\frac{\partial u}{\partial t} = \left[-\frac{\hbar^2}{2m}(\partial_i - (i/\hbar)A_i)^2 + \frac{1}{2I}\big(\widehat{M}_i + \mathfrak{m}IB_i\big)^2 + V\right]u + \frac{2}{\sin\alpha}\frac{\delta Q(q,\theta)}{\delta\psi^*(x,\alpha)}. \quad (33)$$

The corresponding unified field (10) is given by

$$u(x,\alpha,t,q_0,\theta_0) = \psi(x,\alpha,t) - \frac{1}{\sin\alpha\,\psi^*(x,\alpha,t)}\delta\big(x - q(t,q_0,\theta_0)\big)\delta\big(\alpha - \theta(t,q_0,\theta_0)\big). \quad (34)$$

Multiplying (33) and (34) by $\sin\alpha\,u_a^*$ and using (21), the solution of the inhomogeneous equation in the discrete representation is given by $U^a(x,t) = \int u_a^*(\alpha)\,u(x,\alpha,t)d\Omega$ with

$$U^a(x,t,q_0,\theta_0) = \Psi^a(x,t) - \frac{1}{\Psi^{b*}(x,t)v_b^a\big(\theta(t,q_0,\theta_0)\big)}\delta\big(x - q(t,q_0,\theta_0)\big) \quad (35)$$

where $v_b^a(\alpha) = u_b^*(\alpha)/u_a^*(\alpha)$. We thus obtain, in (34) or (35), a unified field for spin $\frac{1}{2}$ similar in form to (2) for spin 0, that is, a solution of the homogeneous equation superposed with a singular soliton (delta function) field modulated by the inverse homogeneous field. In (35), the modulation depends on the rotator basis functions (through the functions $v_b^a(\alpha)$) evaluated along the angular trajectory, and the translational trajectory depends on the initial angle coordinates $\theta_0^r$, as expected from the coupling of the guidance equations (31) and (32). This construction ensures that the function (35) is a 2-spinor (as we prove in the next section), a result that could not have been achieved using just the spinor $\Psi^a$ and the translational trajectory.

As regards the modelling of the corpuscle by the unified field (35), the function $\delta(x-q)$ represents its centre of mass faithfully but the extended character of the body is portrayed only indirectly through the dependence of the modulating functions on the moment of inertia $I$ (for an example, see below).

In general, the corpuscle's space trajectory $q_i(t)$, derived from $v_i$ via (31), does not coincide with an integral curve $q_{Pi}(t)$ of the Pauli velocity built from the homogeneous spinor,

$$v_{Pi} = (\hbar/2mi\Psi^\dagger\Psi)[\Psi^\dagger\partial_i\Psi - (\partial_i\Psi^\dagger)\Psi] - A_i/m. \tag{36}$$

The velocity (36) is used in the causal theory of the discrete Pauli equation (29), and its construction entails a considerable loss of information about the system compared with $v_i$ [5]. The relation between the two velocities is that the Pauli velocity is the mean over the angles of the rotator velocity:

$$v_{Pi}(x) = \int |\psi|^2\, v_i(x,\alpha) d\Omega \bigg/ \int |\psi|^2\, d\Omega. \tag{37}$$

The two species of space trajectory ($q_i$ and $q_{Pi}$) do coincide, however, in an important special case, which in part reproduces the spin 0 treatment: when the wavefunction factorizes as $\psi = \Phi(x,t)\chi(\alpha)$ (we assume $\mathfrak{m}B_i = 0$ here). From (21), we then have $\Psi^a = \Phi(x,t)c^a$, $c^a = $ constant, and, orienting the axes so that $c^a = \delta_1^a$ (spin up), the spin up unified field (35) is given by

$$U^1 = \Phi(x,t) - \frac{\delta(x-q(t,q_0))}{\Phi^*(x,t)}, \qquad U^2 = -\frac{u_2^*(\theta(t,\theta_0))}{u_1^*(\theta(t,\theta_0))}\frac{\delta(x-q(t,q_0))}{\Phi^*(x,t)}. \tag{38}$$

The function $\Phi$ obeys the spin 0 Schrödinger equation and, from (31), the space trajectory is independent of the angle variables, which play no role in $U^1$. This component of the unified field thus obeys the spin 0 inhomogeneous equation. On the other hand, the component $U^2$ is correlated with the independent evolution of the angles. For a free spin up state ($\psi = \Phi u_1$), the solution to (32) is $\theta^1 = \theta_0^1, \theta^2 = \theta_0^2 - \nu t, \theta^3 = \theta_0^3 - \nu t$ with $\nu = \hbar/4I\cos^2(\theta_0^1/2)$ [5], and $U^2$ (indirectly) expresses the extended character of the body through the factor $u_2^*/u_1^*$, which depends on $I$ via the periodic term $e^{i\nu t}$. Similar results are obtained for spin down where the roles of $U^1$ and $U^2$ are interchanged.

It is straightforward to extend the unified theory to any spin value, and, following our previous example [1], to a many-spin system by increasing the range of the coordinate indices.

## 5 Symmetry group of the spin $\frac{1}{2}$ unified theory

### 5.1 Homogeneous field

The Pauli equation (29) is covariant with respect to the continuous symmetries of the 10-parameter Galileo group [12]:

$$t' = t + d, \quad x'_i = a_{ij}x_j - w_i t + c_i,$$
$$\Psi'^a(x',t') = e^{i\chi(x,t)/\hbar} S^a_b \Psi^b(x,t), \quad \chi = m(\mathbf{w}^2 t/2 - a_{ij}w_i x_j), \tag{39}$$
$$A'_i(x',t') = a_{ij}A_j(x,t), \quad B'_i = a_{ij}B_j, \quad V' = V - a_{ij}w_i A_j.$$

Here $d$ (time translation), $c_i$ (space translation), $a_{ij}$ (proper rotation: $a_{ij}a_{ik} = \delta_{jk}$, $\det(a_{ij}) = 1$), $w_i$ (boost) and $S^a_b$ are constants, where $S^a_b \in SU(2)$ with $S^{b*}_a = S^{-1a}_b$ and $a_{ij}\sigma_{jab} = S^{-1c}_a \sigma_{icd} S^d_b$. It suffices to consider infinitesimal rotations with angles $\varepsilon_i, i = 1,2,3$, about the space axes, for which $a_{ij}(\varepsilon) = \delta_{ij} + \varepsilon_k \varepsilon_{ijk}$ and $S^a_b(\varepsilon) = \delta^a_b + \frac{1}{2}\varepsilon_i \sigma^a_{ib}$, where $\varepsilon_{ijk}$ is the antisymmetric symbol. The connection between the transformations of the external fields and Lorentz transformations is explored in [12,13].

To show that the inhomogeneous equation (33) shares the Galileo symmetries, we consider first the covariance of the homogeneous equation (27) by expressing (39) in the angular coordinate representation. We begin with a rotation, for which the generator $S^a_b(\varepsilon)$ acting on discrete indices corresponds to the real operator $\widehat{D}(\varepsilon, \alpha) = 1 + i\varepsilon_i \widehat{M}_i(\alpha)/\hbar$ acting on angle indices. Applied to the basis functions (19), the correspondence becomes

$$\widehat{D}(\varepsilon)u_a(\alpha) = u_b(\alpha)S^b_a(\varepsilon), \tag{40}$$

as may be checked using (22) and (30). To proceed, we observe that the homogeneous equation (27) admits an independent angle symmetry. Consider the following infinitesimal variation of the Euler angles, where the parameter $\eta_i$ need not equal $\varepsilon_i$:

$$\alpha'^r = \widehat{D}(\eta, \alpha)\alpha^r = \alpha^r + \eta_i A^r_i(\alpha). \tag{41}$$

This group of transformations induces the following transformations of the basis functions and angular momentum operators:

$$u_a(\alpha') = \widehat{D}(\eta, \alpha)u_a(\alpha) \tag{42}$$

$$\widehat{M}_i(\alpha') = \widehat{D}(\eta, \alpha)\widehat{M}_i(\alpha)\widehat{D}^{-1}(\eta, \alpha) = a_{ik}(-\eta)\widehat{M}_k(\alpha). \tag{43}$$

These formulas may be proved by replacing $\alpha^r$ with $\alpha'^r$ in (19) and (22), and Taylor expanding. The wavefunction (18) is, therefore, an invariant function: $\psi'(x, \alpha', t) = \widehat{D}(\eta, \alpha)\psi(x, \alpha, t) = \psi(x, \alpha', t)$. We shall show in a moment that (27) is symmetric under (41), as a component of a more general rotational symmetry. The variation (41) evidently corresponds to the identity symmetry of the discrete equation (29). Any symmetry of (27) corresponding to a continuous symmetry of (29) will, therefore, exhibit this angular freedom.

Combining independent rotations of the translational ($x^i$) and rotational ($\alpha^r$) coordinates, the transformed wavefunction is $\psi'(x', \alpha', t') = \Psi'^a(x', t')u_a(\alpha')$. Using (39), (40), (42) and group composition gives $\psi'(x', \alpha', t') = \widehat{D}(\varepsilon + \eta, \alpha)\psi(x, \alpha, t)$. Two cases are of interest: $\eta_i = 0$, for which the Euler angles do not change, and $\eta_i = -\varepsilon_i$, for which the wavefunction is a rotational scalar function on $\mathfrak{M}$: $\psi'(x', \alpha', t') = \psi(x, \alpha, t)$.

To demonstrate the covariance of the homogeneous equation (27) under the combined rotation, consider the equation written with respect to the primed variables. For the magnetic field-spin interaction term we have, from (39) and (43),

$$B'_i(x')\widehat{M}_i(\alpha')\psi'(x', \alpha', t') = a_{ij}(\varepsilon)B_j(x)a_{ik}(-\eta)\widehat{M}_k(\alpha)\widehat{D}(\varepsilon + \eta, \alpha)\psi(x, \alpha, t)$$
$$= a_{kj}(\varepsilon + \eta)B_j\widehat{M}_k\widehat{D}(\varepsilon + \eta)\psi(x, \alpha, t). \qquad (44)$$

Applying the operator $\widehat{D}^{-1}(\varepsilon + \eta)$ and using (43), (44) becomes $a_{kj}(\varepsilon + \eta)a_{ki}(\varepsilon + \eta)\widehat{M}_i B_j \psi = B_i \widehat{M}_i \psi$. Hence, applying $\widehat{D}^{-1}(\varepsilon + \eta)$ to the other terms in the primed equation, we recover (27), which is, therefore, covariant.

Next, we observe that the Euler angles and associated functions are invariant under spacetime translations and boosts so the angular wavefunction just picks up the phase factor $e^{i\chi/\hbar}$ in (39). In sum, the most general transformation of the angular wavefunction corresponding to (39) is

$$\psi'(x', \alpha', t') = e^{i\chi(x,t)/\hbar}\widehat{D}(\varepsilon + \eta, \alpha)\psi(x, \alpha, t). \qquad (45)$$

### 5.2 Inhomogeneous field

It is evident from the unified field solution (34), which is not linear in $u_a$, that the angle transformation (41) is not generally an independent symmetry of the inhomogeneous equation (33), or of the equivalent guidance equations (31) and (32). To establish rotational covariance requires a combination of the transformations parameterized by $\eta_i$ and $\varepsilon_i$, for which $\eta_i = -\varepsilon_i$ in (45). With this assumption, $\psi$ is a rotational scalar so that (45) becomes $\psi'(x', \alpha', t') = e^{i\chi(x,t)/\hbar}\psi(x, \alpha, t)$, and the particle coordinates transform as

$$q'_i(t') = a_{ij}(\varepsilon)q_j(t) - w_i t + c_i, \qquad \theta'^r(t') = \theta^r(t) - \varepsilon_i A^r_i(\theta(t)). \qquad (46)$$

The term $\delta(x' - q') = \delta(x - q)$ since $\det(a_{ij}) = 1$, and $\delta(\alpha' - \theta')/\sin\alpha' = \delta(\alpha - \theta)/\sin\alpha$ since $\sin\alpha' = \sin\alpha[\det(\partial\alpha'/\partial\alpha)]^{-1}$ from (41). The appearance of $1/\psi^*$ in $u$ implies that the latter's two summands have the same phase. Combining these results, we conclude that the unified field transforms like $\psi$:

$$u'(x', \alpha', t', q'_0, \theta'_0) = e^{i\chi(x,t)/\hbar}u(x, \alpha, t, q_0, \theta_0). \qquad (47)$$

Multiplying $u'(x', \alpha', t', q'_0, \theta'_0)$ by $\sin\alpha' u^*_a(\alpha')$, integrating over $\alpha'^r$, and using the result $u^*_a(\theta') = u^*_b(\theta)S^{Tb}_a = S^a_b u^*_b(\theta)$ derived from (40), (42) and (46), we deduce that the field $U^a$ transforms as a 2-spinor:

$$U'^a(x', t', q'_0, \theta'_0) = e^{i\chi(x,t)/\hbar}S^a_b U^b(x, t, q_0, \theta_0). \qquad (48)$$

Having established the transformation laws for the quantities appearing in it, it is easy to show that the inhomogeneous equation (33) is symmetrical with respect to the Galileo group. Form invariance requires that

$$i\hbar \frac{\partial u'}{\partial t'} = \left[ -\frac{\hbar^2}{2m}(\partial'_i - (i/\hbar)A'_i)^2 + \frac{1}{2I}\left(\widehat{M}_i(\alpha') + \mathfrak{m}IB'_i\right)^2 + V' \right] u' + \frac{2}{\sin\alpha'} \frac{\delta Q'(q', \theta')}{\delta \psi'^*(x', \alpha')}. \quad (49)$$

The left-hand side of (49) and the term involving the square brackets are together equal to $e^{i\chi(x,t)/\hbar}$ times the corresponding set of terms in (33). Similarly, from (8) and (15), the primed source term is $e^{i\chi/\hbar}$ times the original term. Hence, (49) is equivalent to (33) and the Galileo covariance of the inhomogeneous equation is proved.

This result may be confirmed by demonstrating the covariance of the equivalent guidance equations (31) and (32), which follows easily using (39), (43), (46), and that $S$ is a rotational scalar. Note that we achieve a more detailed account of covariance than is possible in the discrete formalism, where the velocity, (36) or (37), and the spin vector,

$$s_i(x) = \hbar \Psi^\dagger \sigma_i \Psi / 2 \Psi^\dagger \Psi = \int \psi^* \widehat{M}_k \psi \, d\Omega \Big/ \int |\psi|^2 \, d\Omega, \quad (50)$$

are defined through sums over the spin indices (discrete or angular). Our treatment is, however, in accord with the Galileo covariance of the classical theory, equations (31) and (32) being those expected in the Hamilton-Jacobi formalism for a classical rigid rotator.

## 6 Conclusion

We have extended the unified theory of wave and guidance equations to embrace systems described by the Schrödinger equation in a Riemannian space with an external vector potential. The basic structure found previously (Sect. 1) remains intact in this generalized context: the source term in the inhomogeneous equation (16) depends on the particle microdensity via the quantum potential, and the state function (10), a simple generalization of (2), is correlated uniquely with the de Broglie-Bohm trajectory law (12). Making a suitable choice of metric and position coordinates describing translational and rotational freedoms, we then derived the unified theory for a spin $\frac{1}{2}$ rotator in the angular coordinate representation. The spin theory inherits the property of unique correlation, now between the field (34) and the trajectory equations (31),(32). In the discrete representation (35) of the unified field, the homogeneous component is a 2-spinor, and the centre of mass of its particle component is represented by a solitonic (delta) function that is peaked around the 3-space trajectory and modulated by the homogeneous spinor and a function of the angular variables. This ensures that the unified field as a whole is a 2-spinor.

A deviation in the generalized trajectory law (12) will be reflected in a modification of the solution (10). Following our previous analysis (Sect. 6 in [1]) the modification may be expressed in terms of the Green function for the homogeneous equation. This is particularly pertinent to the spin case. Treating the Pauli theory as the non-relativistic residue of the Dirac theory, it has been shown [14-17] that the Pauli velocity (36) must acquire an additional spin-dependent term,

$$v_{si} = (1/m\Psi^\dagger\Psi)\varepsilon_{ijk}\partial_j(\Psi^\dagger\Psi s_k), \tag{51}$$

so that the total velocity is $v_{Pi} + v_{si}$. The term (51) plays a significant role in several contexts, such as two-slit interference [16] and arrival times [18-20]. It corresponds to the addition of the expression $\text{Re}\big[(1/m|\psi|^2)\varepsilon_{ijk}\partial_j(\psi^*\widehat{M}_k\psi)\big]$ to the right-hand side of the velocity (31) (cf. (37)). This supplement is consistent in that it leaves invariant the continuity equation for $|\psi|^2$ implied by (27), and preserves Galileo covariance. It is the subject of further enquiry to situate the associated adjustment to the solution (34) within a relativistic unified theory.